\documentclass{paper}
\usepackage{graphicx}
\usepackage{psfrag}
\usepackage{amsmath}
\usepackage{cite}
\usepackage{hyperref}

\title{Indifferents as an interface between Contra and Pro\thanks{for the 4-th Polish Seminar on Econo- and Sociophysics, 2009/05/07-09, Rzesz\'ow, Poland}}
\author{Krzysztof Malarz~\thanks{\tt malarz@agh.edu.pl} and Krzysztof Ku{\l}akowski~\thanks{\tt kulakowski@novell.ftj.agh.edu.pl}}
\institution{
AGH University of Science and Technology,
Faculty of Physics and Applied Computer Science,
al. Mickiewicza 30, PL-30059 Krak\'ow, Poland
}
\date{\today}
\begin{document}
\maketitle

 \begin{abstract}
In most sociophysical simulations on public opinion, only two opinions are allowed: Pro and Contra. However, in all political elections many people do not vote. Here we analyse two models of dynamics of public opinion, taking into account Indifferent voters: {\it i)} the Sznajd model with symmetry Pro-Contra, {\it ii)} the outflow one move voter model, where Contra's are converted to Indifferent by their Pro neighbours.
Our results on the Sznajd model are in an overall agreement with the results of the mean field approach and with those known from the initial model formulation. The simulation on the voter model shows that an amount of Contra's who remain after convertion depends on the network topology.
 \end{abstract}
 
 \noindent
 
 {\em PACS(2008) numbers:}  
89.65.-s, 
87.23.Ge, 
89.65.Ef, 
89.75.-k, 
89.75.Fb  
 
 \noindent
 

 \bigskip
 \section{Introduction}

Attempts of statistical physicists to deal with sociological problems is an occurrence relatively 
new, but the question on the specific character of social sciences lasts for the whole 
history of modern sociology \cite{szacki}. It seems that the gap between sociophysics and mathematical sociology is 
more narrow than the one between empiricistic and hermeneutical sociologies. From this point of view, 
sociophysics can be seen as an overlap of physics and sociology. Mathematical modeling of public opinion, which 
are of interest here, covers quite a large part of this area. A recent review can be found in Ref. \cite{santo}. \\

As it is known, numerous statistical data on opinions include simple {\it Don't know} as a third possible answer \cite{cbos}. Also, indifferents are the target of most electoral campaigns. This means that once the mere existence of this third option is disregarded, it is hard to understand 
the time dynamics of public opinion on any issue. On the other hand, it seems that once two options are symmetrical, an intermediate option is 
necessary to give a complete picture. This rough notion on symmetry is based on two examples. The case of election with two candidates 
is symmetrical and the indifferent option is natural. The case of support for an ecological protest is not symmetrical, because 
the indifferent option means `no support'.\\

Motivation of our work comes from the fact that in the current models of public opinion, neutral or indifferent opinions are usually neglected. Opinions either vary continuously between Pro (P) and Contra (C), or --- in discrete version --- they are limited to these two cases, P or C. The models of Hegselmann--Krauze \cite{hegs} and of Deffuant-Weisbuch \cite{deff} belong to the first cathegory, while the Sznajd model \cite{sznajd,sz1,sz2,sz3,sz4,sz5,sz6,sz7,sz8,sz9,sz10,sz11} falls into the second one, with \cite{sz11} as an exception. In some papers, third option is considered as third political party \cite{sza} or third language \cite{hadzi}, or three states are completely different in their nature \cite{pelto}. The option of the symmetry Pro-Contra with the neutral state as intermediate one is considered rather rarely. 
We can refer to the text of Yang {\em et al.} \cite{yang}, where the set of states of a network 
node includes $0$ as the third option apart from $\pm 1$. In this paper, the time dynamics of opinions is ruled by the same 
law as interacting spins in the presence of magnetic field. Further, in papers by He {\em et al.} \cite{hesun} and by Dall'Asta 
and Galla \cite{dalla} the voter model with the third state has been simulated on two-dimensional lattices.  \\

We are going to consider this third option of Indifferent (I), {\em i.e.} neutral state in two models of public opinion on random networks. The first application is the Sznajd model with the symmetry Pro-Contra preserved. The second application is the outflow voter model, where this symmetry is broken. 
The purpose of the latter is to refer to the sociological concept of the spiral of silence \cite{noelle}. The idea of spiral of silence 
means that there is a positive feedback between the social acceptance of a given opinion and the willingness of its adherents to state it in 
open way. Then, the minority opinion gets less and less support and finally nobody admits to it. Theory of spiral of silence was 
formulated by Elisabeth Noelle--Neumann in her search for origins of the weakness of the Anti-Nazi opposition in Germany. This 
historical case is of extremal importance, but less famous examples are more frequent. In 2007, an information in Polish TV about 
minister Kaczmarek influenced the poll outcome on support for two main parties, because adherents of the Civic Platform preferred
to answer {\it Don't know} after this embarrassing transmission \cite{pacewicz}. This variation, clearly exceeding statistical errors, 
vanished after a few days.\\

The simulation is performed for the Erd{\H o}s--R{\'e}nyi network with various mean nodes degrees $\langle k\rangle$ and, occasionally, for the growing scale-free and exponential trees.
In two subsequent sections we explain the formulation and show the results of the symmetric and antisymmetric cases, respectively.
Last section is devoted to discussion.

\section{The Sznajd model}

In its original formulation \cite{sznajd}, the Sznajd model postulated two states, say P and C (Pro and Contra). The rule of time evolution was as follows. A pair of neighboring nodes, randomly selected, was checked if the nodes were in the same states. If this was the case, this state (P or C)
was assigned also to all neighbours of each node of the pair. The basic result was that after a sufficiently long time, the whole system fell into one of the two states, P or C. \\

Our formulation which includes zero state ({\it Don't know}) and preserves symmetry P-C is defined as follows. A pair of neighboring nodes ($i$ and $j$), if they 
are in the same state P ($S_i=S_j=+1$) or C ($S_i=S_j=-1$), share this state with their neighbours $n$. If the paired nodes are in opposite states (P-C or C-P), all neighbours of these nodes become neutral (state $S=0$). If any of two nodes is in the neutral state ($S_i=0$ or $S_j=0$), nothing changes:

\begin{equation}
S_n(t+1)=
\begin{cases}
S_i(t)=S_j(t) & \text{ if } S_i(t)S_j(t)=1,  \\
0             & \text{ if } S_i(t)S_j(t)=-1, \\
S_n(t)        & \text{ if } S_i(t)S_j(t)=0.  \\
\end{cases}
\end{equation}

\begin{figure}
\psfrag{delta=0.1}{(a) $\delta=0.1$}
\psfrag{delta=0.15}{(b) $\delta=0.15$}
\psfrag{delta=0.2}{(c) $\delta=0.2$}
\psfrag{delta=0.25}{(d) $\delta=0.25$}
\psfrag{delta=0.30}{(e) $\delta=0.3$}
\psfrag{delta=0.35}{(f) $\delta=0.35$}
\psfrag{delta=0.40}{(g) $\delta=0.4$}
\psfrag{delta=0.45}{(h) $\delta=0.45$}
\psfrag{densities of R, G, 0}{$\rho_P, \rho_C, \rho_I$}
\includegraphics[width=0.45\textwidth]{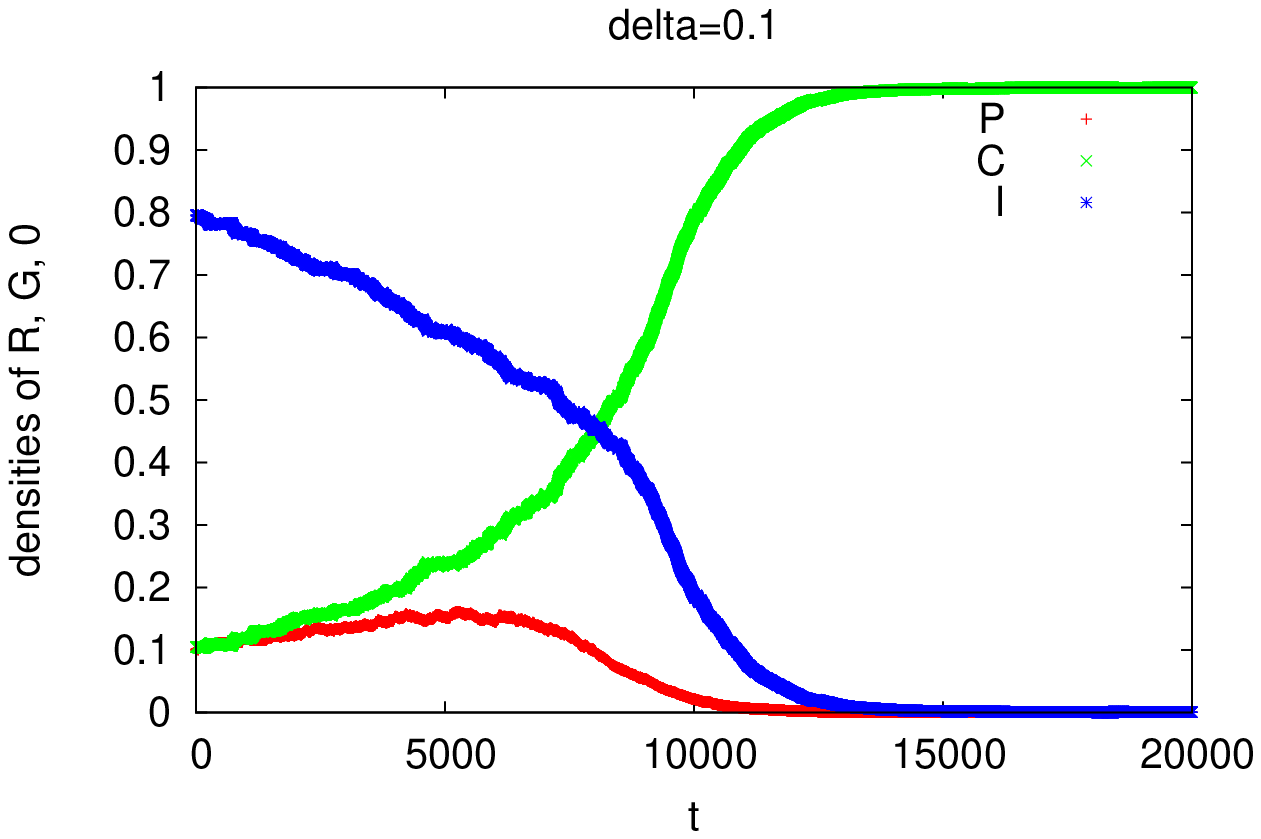} 
\includegraphics[width=0.45\textwidth]{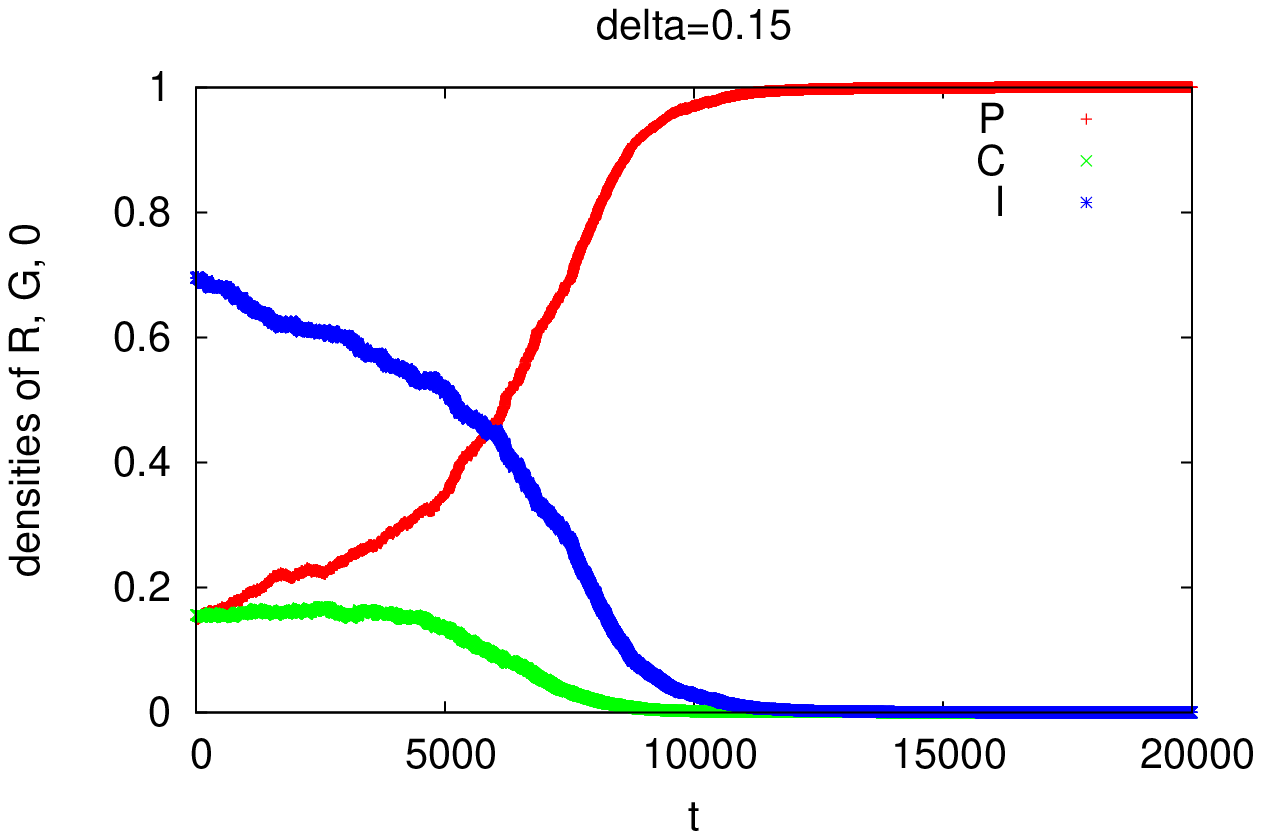}\\
\includegraphics[width=0.45\textwidth]{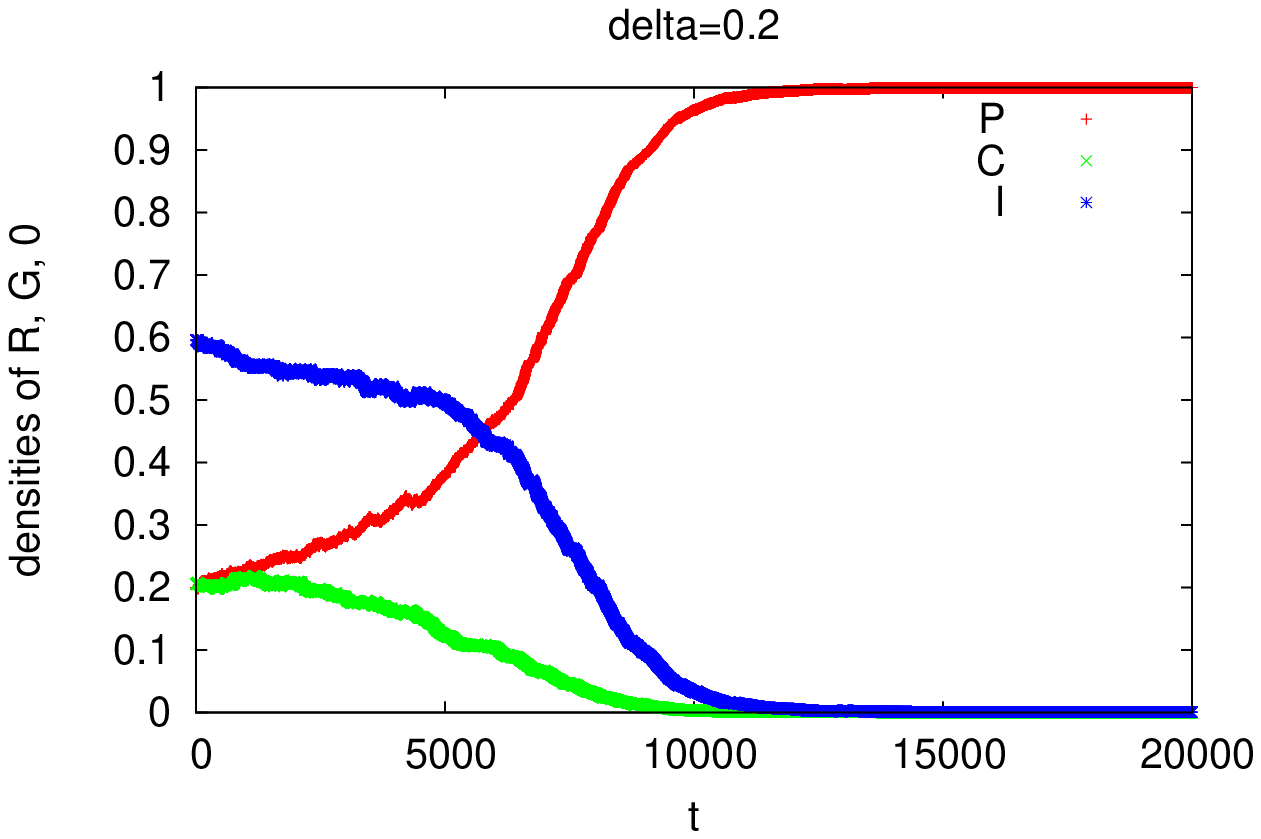} 
\includegraphics[width=0.45\textwidth]{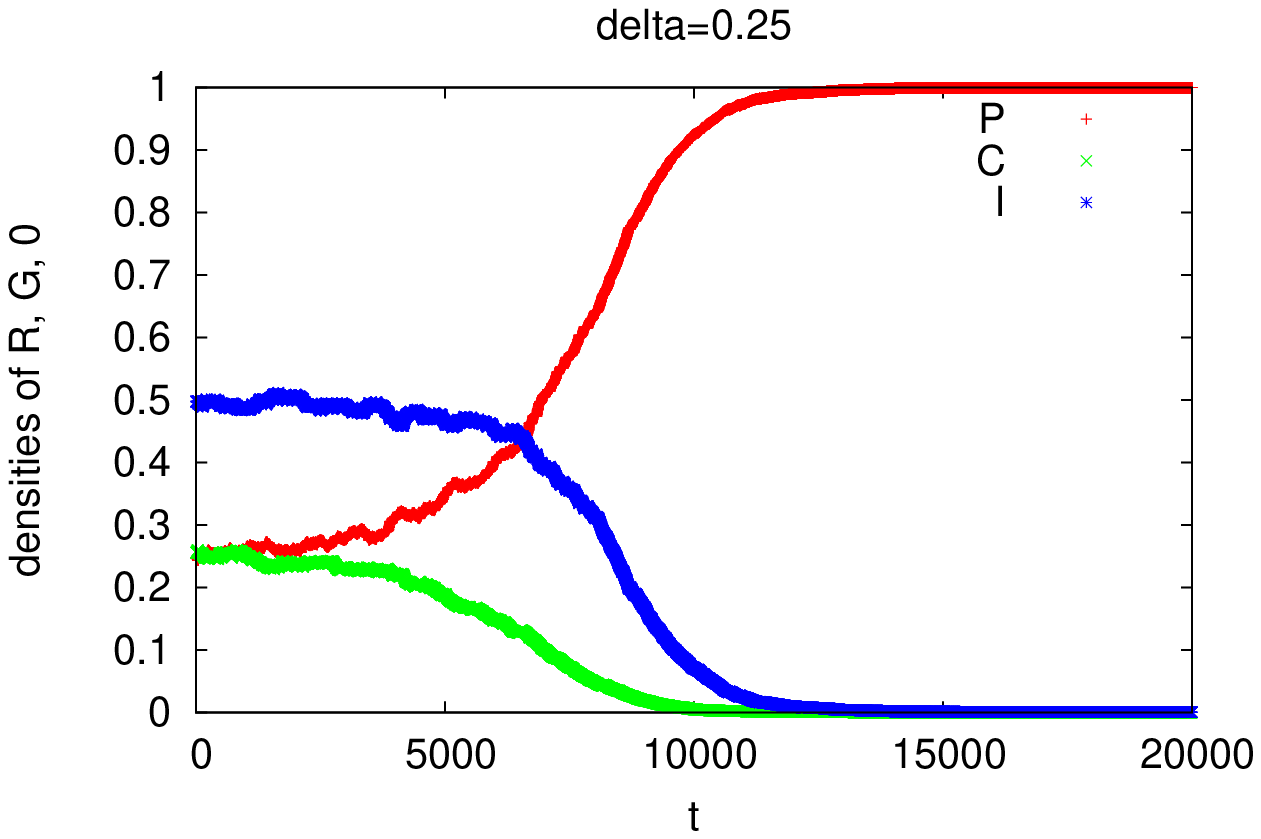}\\
\includegraphics[width=0.45\textwidth]{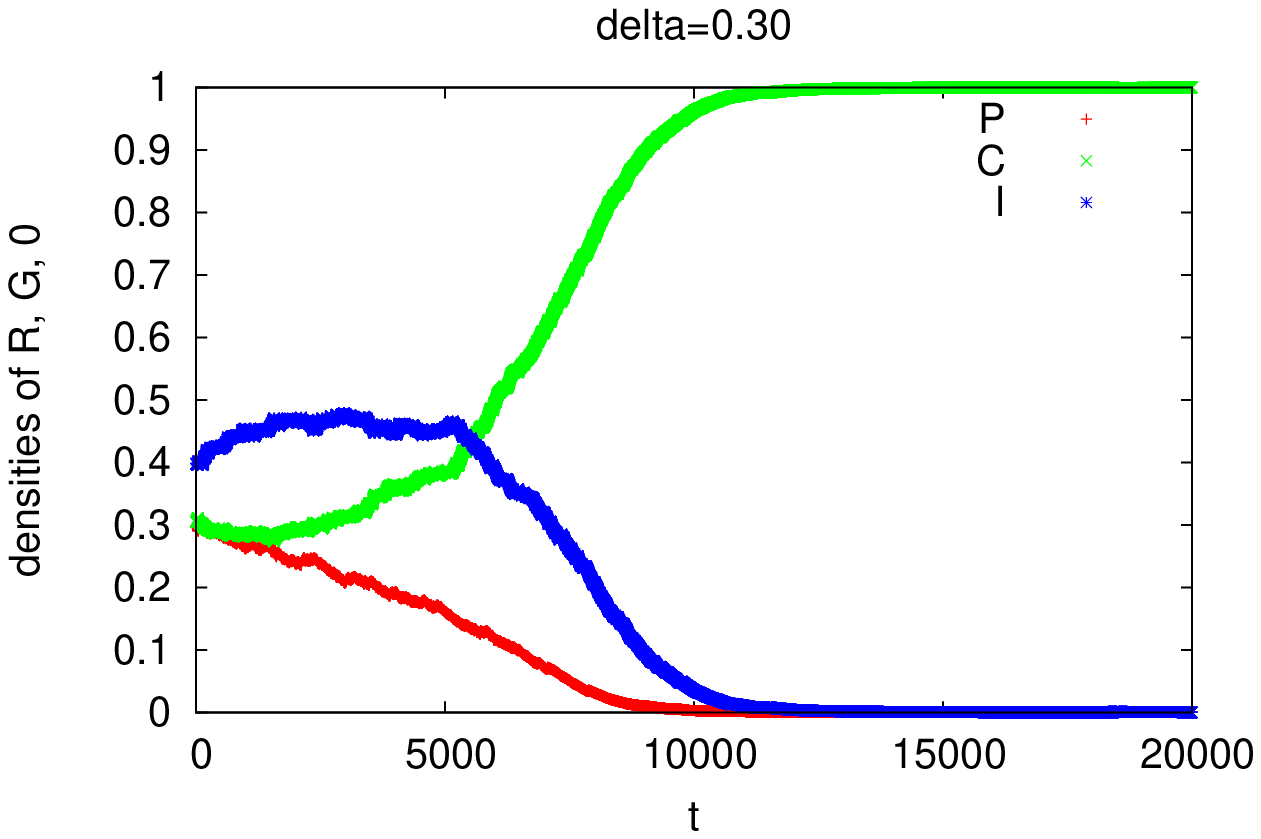} 
\includegraphics[width=0.45\textwidth]{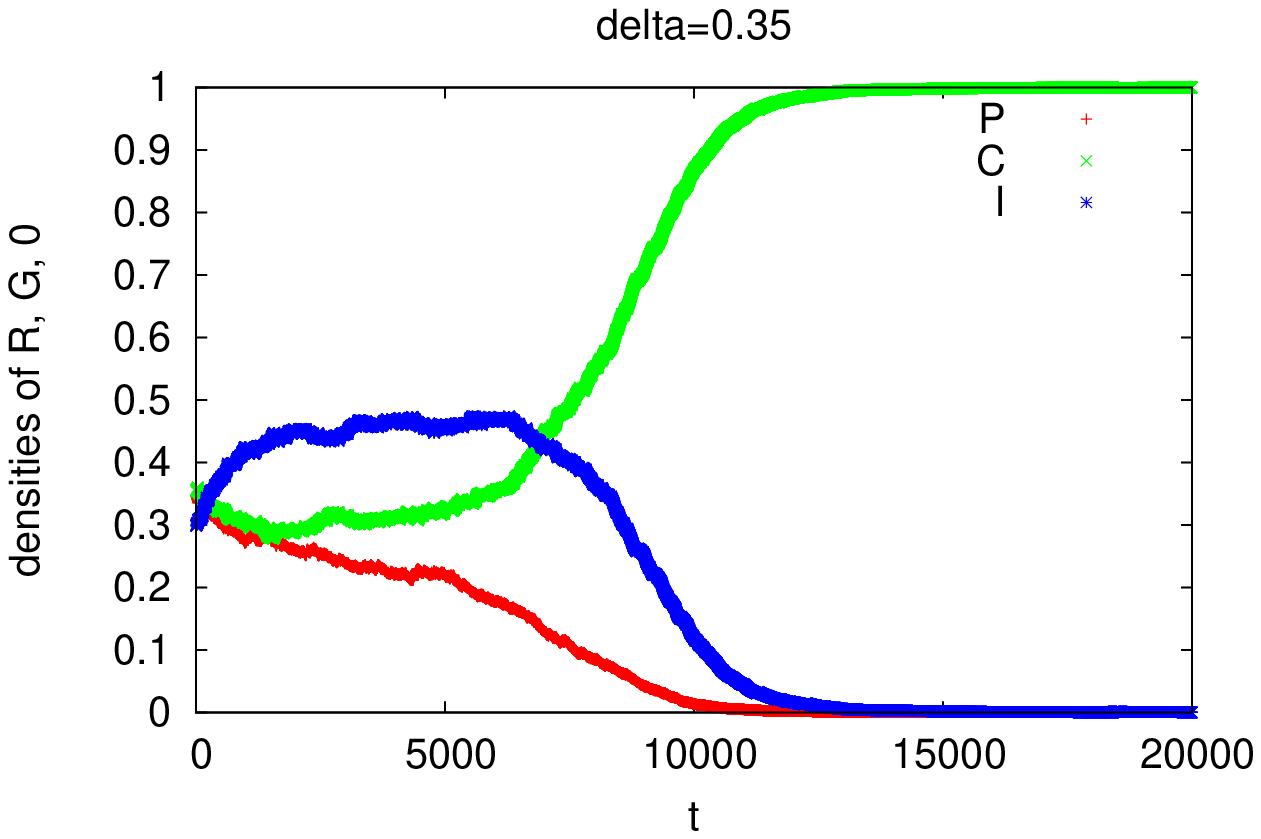}\\
\includegraphics[width=0.45\textwidth]{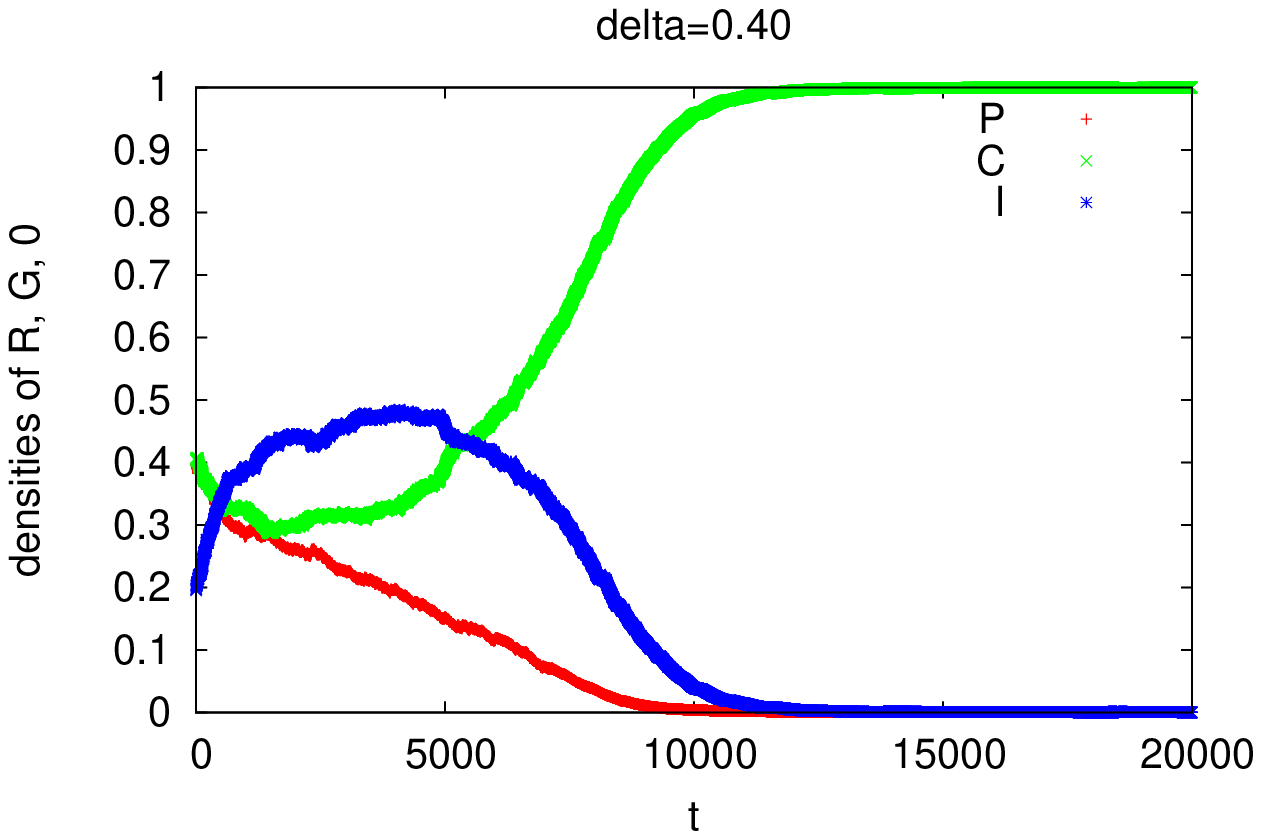} 
\includegraphics[width=0.45\textwidth]{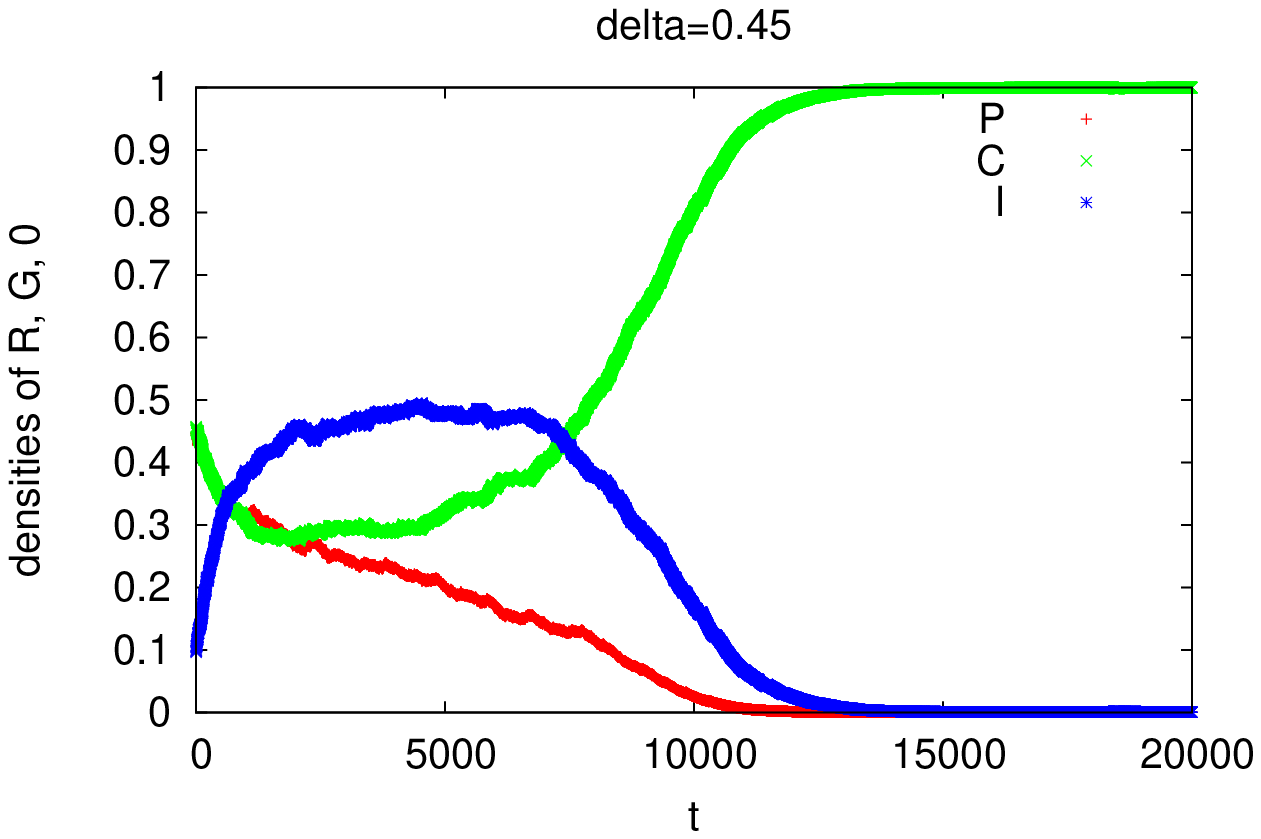} 
\caption{Time evolution of the densities $\rho$ of P, C and I for various initial concentration $\delta$ of P and C.
$N=10^4$, $\langle k\rangle=10$.}
\label{timevol}
\end{figure}

Although this fomulation seems to be less restrictive than the original one, our numerical results indicate that the result is basically the same. In all investigated cases, the final state of the network is either all-P ($\sum_i S_i=+N$) or all-C ($\sum_i S_i=-N$) for the whole network. Examples of the time dependence of the amount of P and C are shown in Fig. \ref{timevol} for different initial states, {\em i.e.} for various values of $0<\delta<1/2$ parameter.
Initially, a fraction $\delta$ of nodes stay in state P ($S_i=+1$) and C ($S_i=-1$) while $(1-2\delta)N$ actors is neutral ($S_i=0$).

This result can be compared with the mean-field calculation, where all local configurations of the network are approximated by an assumption of perfect mixing. In this case, the time dependence of the concentration of Pro ($x(t)$) and Contra ($y(t)$) can be written as
\begin{subequations}
\begin{equation}
\dot x = x^2(1-x) - xy^2 - \alpha x^2y,
\label{eq-mfa-x}
\end{equation}
\begin{equation}
\dot y = y^2(1-y) - yx^2 - \alpha y^2x,
\label{eq-mfa-y}
\end{equation}
\label{eq-mfa}
\end{subequations}
and the concentration of nodes in the neutral state I is $z=1-x-y$.
Equations \eqref{eq-mfa} preserves the P-C symmetry.
On the right-hand-side of Eq. \eqref{eq-mfa-x}, the first term describes the process where two nodes in state P ($x^2$) convert a node in the `non-P' state ($1-x$) to the P state.
Next term is responsible to a conversion of P ($x$) into non-P by two C ($y^2$).
The last term is due to the conversion of P ($x$) to I by a P-C pair ($xy$).
We give in parenthesis terms which the appropriate rates are proportional to.
In principle, the rate of the last process in not necessarily equal to the others two.
To mark the possible difference of these rates, we introduce a constant $\alpha$.\\

According to the known procedure \cite{glen}, we find the fixed points $(x^*,y^*)$ and determine their stability; only stable fixed points are meaningful.
The stability condition is that both eigenvalues of the Jacobian, calculated at a given fixed point, are negative.
Here we get four fixed points.
First one is $(0,0)$ and its stability is marginal, as both eigenvalues are equal to zero.
However, this fixed point is not interesting for us, as it corresponds to neutral agents at all nodes of the system.
There are also two fixed points $(0,1)$ and $(1,0)$ with eigenvalues $(-1,-1)$ for any value of $\alpha$; these fixed points are always stable.
The last fixed point is $x^*=y^*=\beta$, where $\beta=1/(2+\alpha)$, and its eigenvalues are $\pm\beta$; this is unstable.
As we see, in this case the mean-field solution agrees with the results of the numerical calculations.

\section{The voter model}

On the contrary to the standard formulation of the voter model \cite{santo,liggett}, here we adopt the outflow version of the voter model, where nodes influence the state of all their neighbours. This difference makes sense in random networks, where neighboring nodes differ in their degrees \cite{reds}.
The outflow behaviour, in the spirit of the Sznajd model, is motivated by the analogy with the effect of spiral of silence.
Namely, we imagine that one of two political parties (Pro) gets the power and bans the other party (Contra).
Then, each member of Pro denounces his/her neighbours who belonged to Contra, converting them by force to the neutral state.
Suppose that the initial state concentration is again $\delta$, $1-2\delta$, $\delta$ for P, I (Indifferents) and C, respectively.
How the final amount of Contras depend on $\delta$?
In other words, how large the opposition, now illegal, can be?

\begin{figure}
\psfrag{varphi}{$\varphi$}
\psfrag{delta}{$\delta$}
\psfrag{N}{$N$}
\includegraphics[scale=0.85]{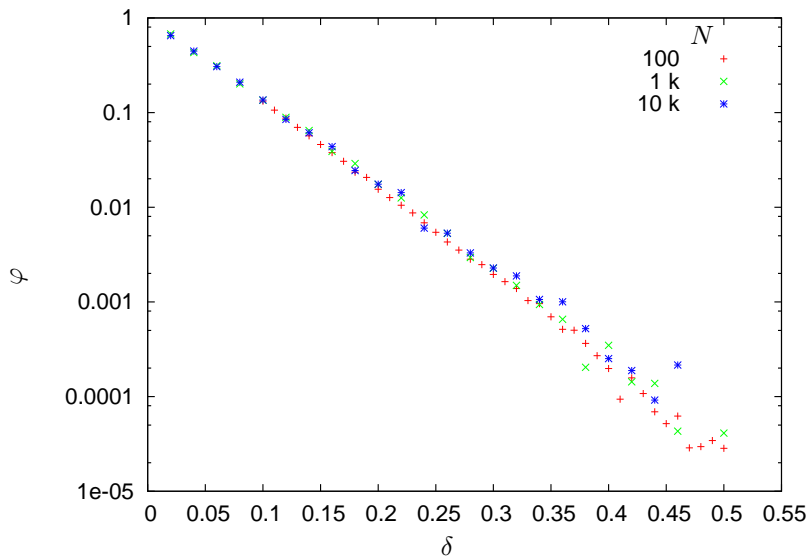}
\caption{Quotient $\varphi$ of final/initial number of C for various system sizes $N$.
Initially, the fractions of P and C among $N$ agents is equal to $\delta$.
The mean nodes degree is $\langle k\rangle=20$.
The results are averaged over $N_{\text{run}}=5000$, 50 and 5 networks realizations for $N=100$, $10^3$ and $10^4$,  respectively.}
\label{varphi-vs-N}
\end{figure}

The simulation --- a strong variant of the spiral of silence --- is designed as to fit this construction.
At the initial state, opinions are assigned randomly to the nodes with probabilities $\delta, 1-2\delta,\delta$ for P, I and C, respectively.
In one step of simulation, all C which have P as neighbours are converted to I.
In Fig. \ref{varphi-vs-N} we show the proportion $\varphi$ of the final amount of C to the initial one, as dependent on $\delta$.
The results can be fitted with the curve
\begin{equation}
\varphi=10^{-\gamma \delta \langle k\rangle},
\end{equation}
where $\gamma \approx 0.4$. The curve obtained for $\langle k\rangle=2$ coincides with the results obtained for the preferentail and randomly growing tree, where $\langle k\rangle$ is also two.
This suggests, that the network topology could be of minor importance.\\

In these calculations, the agents moves are performed only once. Therefore, there is no direct path from the numerical to continuous dynamics. Still for the completeness we refer to the mean field population dynamics, which can be written as 

\begin{subequations}
\label{eq-4}
\begin{equation}
\dot x=x(z-\alpha y)
\end{equation}
\begin{equation}
\dot y=y(z-\alpha x)
\end{equation}
\end{subequations}
and, as before, $z=1-x-y$. These equations are constructed with the following processes in mind: $PI \to PP$, $CI \to CC$, $PC \to II$. The rate of two 
first processes is chosen to be $\alpha $. As before, we get $x=1$ or $y=1$ as the only stable points. It is worthwhile to note that this time evolution
is slightly different from the Krapivsky formulation of the catalytic processes \cite{krapi}, as there atoms P and C arise spontaneously. In the latter formulation we should write $\dot x=z-\alpha xy$, $\dot y=z-\alpha xy$ and $z=1-x-y$. As we see, $x-y$ is a constant of motion. At the only solution 
P and C coexist, on the contrary to the `sociological' formulation in Eqs. \eqref{eq-4}. \\

\begin{figure}
\psfrag{avek}{$\langle k\rangle$}
\psfrag{varphi}{$\varphi$}
\psfrag{delta}{$\delta$}
\includegraphics[scale=0.85]{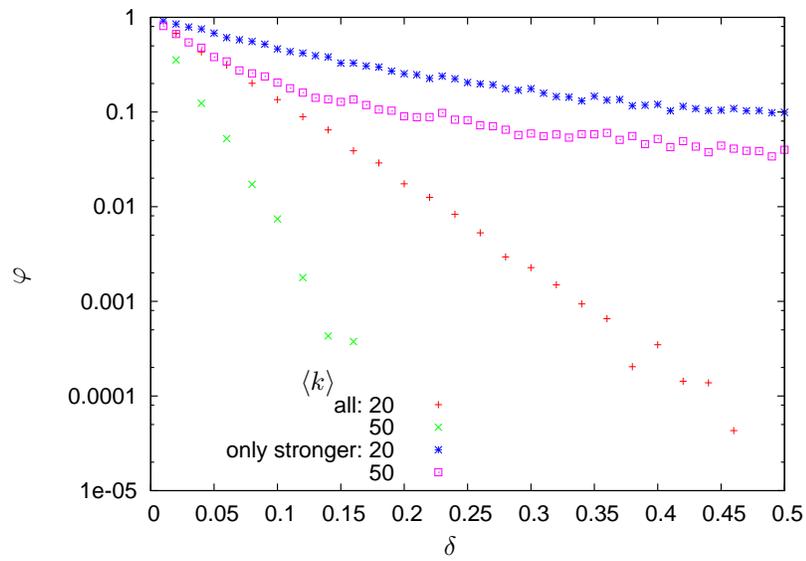}
\caption{C's are neutralized to I's only when at least one of its P's neighbours is `stronger' (stars and squares).
For comparizon the results when each P in C's neigbhbourhood is able to neutralize it are included as well (pluses and crosses).}
\label{fens4_1_4}
\end{figure}

Next, the numerical simulation is repeated in some milder variant, where the action of P against C depends on their degrees. Namely, P is allowed to convert C into I only if P has more neighbours than C. This calculation is motivated by the interpretation of the degree in social networks as an index of prestige \cite{wass}. Then, in this mild scenario only more connected agent can influence the opinion of others. As we see, two meanings of the term `connected',
the topological one and the social one harmonize. The results of this version of simulation shown in Fig. \ref{fens4_1_4} indicate that indeed, the ratio $\varphi$ decreases with $\delta$ clearly slower, than exponentially.\\

\begin{figure}
\psfrag{avek}{$\langle k\rangle$}
\psfrag{varphi}{$\varphi$}
\psfrag{delta}{$\delta$}
\includegraphics[scale=0.85]{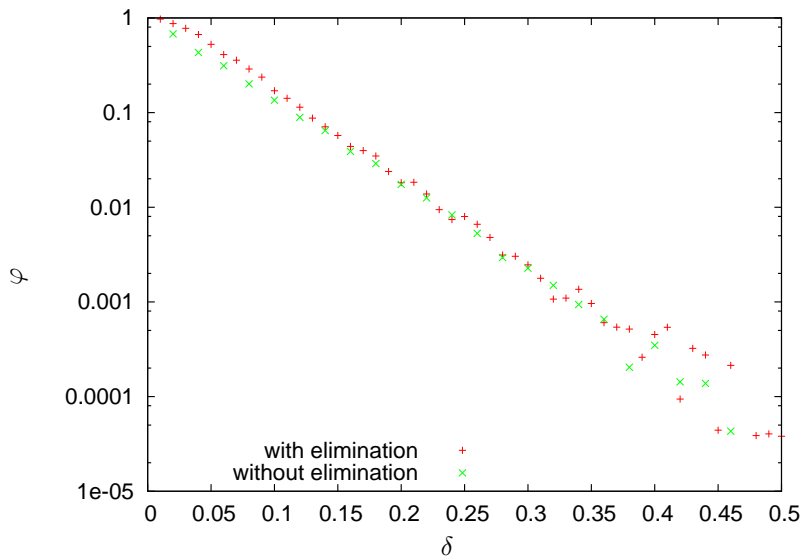}
\caption{(a) P’s without contact with other P’s are eliminated.
(b) Only those P’s are eliminated which have at least one neighbour C}
\label{fens4_4}
\end{figure}

The role of degree can be investigated yet in another variant of calculation. Let us suppose that just before the opinion C is banned, C are able to convert to I all P who have no links to other P's. This modification is applied to the strong variant of the spiral of silence. The motivation of this variant comes from the question, if it is desirable for the opposition C --- just about to be banned --- to neutralize those P who are isolated, by force if necessary. We are glad to announce that this method seems not to be fruitful. Two curves in Fig. \ref{fens4_4} --- with and without this extra neutralization --- almost coincide. Small differences appear only in the range of small values of $\delta$, where the consequences of the whole ban are less severe.\\

\begin{figure}
\psfrag{avek}{$\langle k\rangle$}
\psfrag{psi}{$\psi$}
\psfrag{delta}{$\delta$}
\includegraphics[scale=0.85]{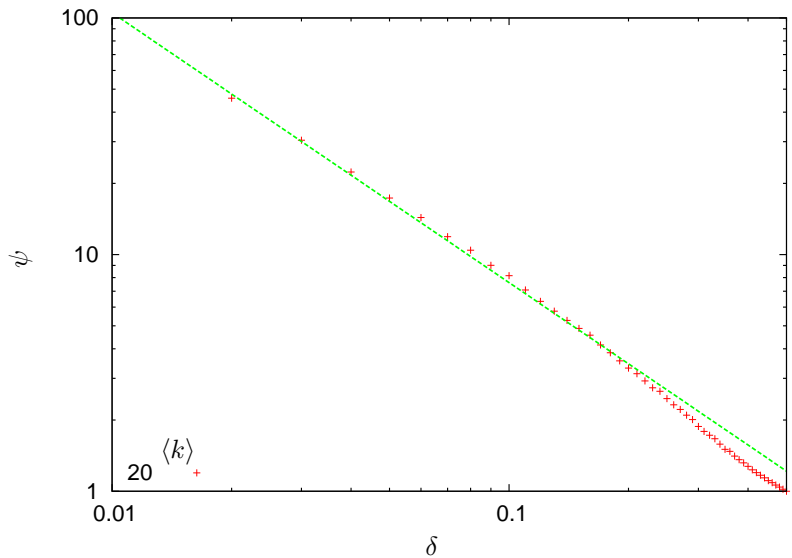}
\caption{All undecided I are forced to vote for P, if they have more neighbors P than C.
$\psi\propto\delta^{-1.14}$ for average node degree $\langle k\rangle =20$.}
\label{fens4_3}
\end{figure}

In the last variant of our simulation the conversion is not from C to I, but from I to P, with two additional conditions. First is that a neutral node I
is converted only if it has more neighbours in the state P than in the state C. Second limitation is that those converted do not convert further.
This calculation is an attempt to refer to the asymmetric scenario, when the ruling opinion is imposed forcefully to those who are neutral or indifferent. History of Central Europe in 1944-49 brings examples, where the percentages of supporters of new power were quite large. In Fig. \ref{fens4_3} we show the rate $\phi$ of those converted to the initial amount of adherents of P. The obtained data can be rougly fitted as $\psi\propto\delta^{-\rho}$ with $\rho\approx 1.14$.
This parameterization reveals, that the smaller is the amount of initial Pro's, the more surprising can be the majority collected in the above mentioned way.\\

\section{Discussion}

We presented a set of simulations motivated by various social and/or historical contexts. The new element of this work is the possibility of the indifferent state in the set of opinions.
As we argued in the Introduction, this modification of the existing models of public opinion is justified by options which are allowed in polls.
This link --- from real data to the model --- evades at least some aspects of the discussion on the relation between sociological phenomena and their measurements \cite{may}.\\

Our results on the Sznajd model are in an overall agreement with the results of the mean field approach and with those known from the initial model formulation \cite{sznajd}. The mechanism of the opinion propagation by creating its new supporters, who propagate it further, is not influenced by the amount of neutral nodes.\\

In the one move voter model neutral nodes can form an environment where the opponents to the option in power can be preserved at least to some extent.
Here, the comparison of the simulation results with those of the mean field approach indicates that the results are sensitive to the network topology.
Networks are defined without space; they have no boundaries but they have peripheries. In the one move voter model the converted nodes do not propagate the ruling opinion. Agents being Contra can then be preserved in the network peripheries, at the low connected nodes.\\

To conclude, the third indifferent option between the dichotomy `Pro-Contra' is an ingredient which shifts mathematical models closer to the results of sociological polls. With this modification, the spectrum of applications of the models gets wider. In our opinion, conversions from Pro to Contra are 
rather rare. When modeling, more attention should be payed to the boundary between Pro and Indifferent and the one between Indifferent and Contra.
Our results indicate, that each of these boundaries is governed by its own dynamics.

\section*{Acknowledgements}
This work was partially supported from the AGH-UST project 11.11.220.01.
The machine time on SGI Altix 3700 in ACK\---CY\-F\-RO\-NET\---AGH is financed by the Polish Ministry of Science and Higher Education under grant No. MEiN/\-SGI3700/\-AGH/\-024/\-2006.

\end{document}